\documentclass[preprints,article,submit,pdftex,moreauthors]{Definitions/mdpi} 
\usepackage{caption}

\firstpage{1} 
\pubvolume{1}
\issuenum{1}
\articlenumber{0}
\pubyear{2023}
\copyrightyear{2023}
\datereceived{ } 
\daterevised{ } 
\dateaccepted{ } 
\datepublished{ } 
\hreflink{https://doi.org/} 

\Title{Pedestrian Accessible Infrastructure Inventory: Assessing Zero-Shot Segmentation on Multi-Mode Geospatial Data for All Pedestrian Types}

\TitleCitation{Pedestrian Accessible Infrastructure Inventory: Assessing Zero-Shot Segmentation on Multi-Mode Geospatial Data for All Pedestrian Types}

\Author{Jiahao Xia $^{1, \ddagger}$, Gavin Gong $^{2, \ddagger}$, Jiawei Liu $^{3, \ddagger}$, Zhigang Zhu $^{4}$ and Hao Tang $^{5}$*}

\AuthorNames{Jiahao Xia, Gavin Gong, Jiawei Liu, Zhigang Zhu and Hao Tang}

\AuthorCitation{Xia, J.; Gong, G.; Liu, J., Zhu, Z. and Tang, H.}

\address{%
$^{1}$ \quad Department of Civil and Environmental Engineering, Rutgers, The State University of New Jersey; jx198@soe.rutgers.edu\\
$^{2}$ \quad East Brunswick High School; gonggavineb@gmail.com\\
$^{3}$ \quad Department of Computer Science, The CUNY Graduate Center; jliu9@gradcenter.cuny.edu\\
$^{4}$ \quad Department of Computer Science, The CUNY City College and The CUNY Graduate Center; zzhu@ccny.cuny.edu\\
$^{5}$ \quad Department of Computer Information Systems, The CUNY BMCC and The CUNY Graduate Center; htang@bmcc.cuny.edu
}
\corres{Correspondence: htang@bmcc.cuny.edu}

\secondnote{These authors contributed equally to this work.}

\abstract{In this paper, a Segment Anything Model (SAM)-based pedestrian infrastructure segmentation workflow is designed and optimized, which is capable of efficiently processing multi-sourced geospatial data including LiDAR data and satellite imagery data. We used an expanded definition of pedestrian infrastructure inventory which goes beyond the traditional transportation elements to include street furniture objects that are important for accessibility but are often omitted from the traditional definition. Our contributions lie in producing the necessary knowledge to answer the following two questions. First, which data representation can facilitate zero-shot segmentation of infrastructure objects with SAM? Second, how well does the SAM-based method perform on segmenting pedestrian infrastructure objects? Our findings indicate that street view images generated from mobile LiDAR point cloud data, when paired along with satellite imagery data, can work efficiently with SAM to create a scalable pedestrian infrastructure inventory approach with immediate benefits to GIS professionals, city managers, transportation owners, and walkers, especially those with travel-limiting disabilities, such as individuals who are blind, have low vision, or experience mobility disabilities.}

\begin{document}

\nolinenumbers

\setcounter{section}{-1} 



\section{Introduction}

Walkability is perhaps the biggest defining characteristic of a healthy and equitable community, as recent literature points to numerous benefits associated with communities with greater walkability \cite{lo2009walkability}. These benefits include, but are not limited to, the reduction of obesity and cardiovascular illness, improved physical well-being, and even a reduction of depression symptoms \cite{boulch2018snapnet}. This is especially true for those with travel-limiting disabilities, such as individuals who are blind, have low vision, or experience mobility disabilities. Historically, the development of large urban centers has been biased towards serving a car-dominant culture, inevitably sacrificing walkability. Increased congestion and air pollution in urban centers, and the undermining of walking as an enjoyable journey in these communities. In our constantly changing climate, the unprecedented number of natural events such as heatwaves and wildfires is also impacting the walkability of communities. It appears that we are being caught in an irreversible downward spiral, a trend where growing car-centric designs and industries are encouraging driving over walking. In fact, these carbon-based automobile industries represent one of the biggest causes of the decline of walkability as they also identify as one of the largest stimuli of climate change, a phenomenon which increases the frequency of extreme weather events and dampens the public’s desire to walk \cite{azmi2012implications}. To combat this downward trend, we need to make significant changes to pedestrian infrastructures in communities so that they can provide convenient, safe, and enjoyable walking experiences for everyone, especially those with visual and mobility disabilities.  

Traditionally, the term ‘pedestrian infrastructure’ refers to a set of transportation elements such as sidewalks, trails, crosswalks, and intersection designs. Expanding this definition with accessibility features allows us to include amenities along the roads, public arts, and wayfinding services to make walking a truly enjoyable experience for all, including people with disabilities. Further expansion of this definition allows for the inclusion of variations of street landscapes, such as tree canopies and street furniture, to address thermal comfort and physical fatigue \cite{azmi2012implications}. These expanded definitions are closely tied to the aspiration to serve all pedestrian types, especially people facing challenges such as those with visual and mobility disabilities, in rehabilitation, or who are elderly \cite{gamache2019mapping}. One major result of this aspiration is an increasing demand for the digitization of current urban environments with a greater level of details so that the state of the pedestrian infrastructure in a community can be assessed with higher confidence.

\textit{Pedestrian infrastructure inventory} is a common means utilized by transportation agencies to assess the state of pedestrian infrastructure and to determine maintenance and improvement needs. There is already a rich body of literature on various methods used for creating pedestrian infrastructure inventory. A significant portion of the work in this domain relied on field observations \cite{frackelton2013measuring, cahen4100935municipal} and street-view photo based analysis \cite{kang2021developing}. Remote sensing technologies enabled the utilization of satellite images and point cloud data for digitizing pedestrian infrastructure \cite{hosseini2021sidewalk,hosseini2022towards, hosseini2023mapping, luo2019developing, senlet2012segmentation, ai2016automated, hou2020network, horvath2021real}. The rise of social media also added an additional dimension of data collection through crowdsourcing and volunteered geographic information (VGI) \cite{omar2022crowdsourcing, erraguntla2017mobile}.  However, it is important to recognize that even with these notable advancements, none of the data collection methods cited above can truly create a complete picture of the pedestrian infrastructure in a community. Particularly when adopting and defining the term ‘pedestrian infrastructure’ in a broader state, it becomes quite apparent that the strengths of each inventory method are balanced out by their weaknesses. This recognition has led to studies exploring fusion of multiple types of data for pedestrian infrastructure inventory \cite{luaces2020accessible, huang2202multi, ning2022sidewalk, hara2014tohme}. 

With the increasing amount of data that can be collected on pedestrian infrastructures, computer vision and machine learning methods have become mainstream research due to the challenge of manually processing these large datasets. Despite numerous proposed computer vision and machine learning model for pedestrian infrastructure inventory, in particular for accessible sidewalk inventory, the new categories of pedestrian infrastructure being considered by equity-minded researchers present significant challenges to existing pedestrian inventory practices and research, a consequence of the unlikelihood that data from single sources and AI models trained for extracting specific features would be able to provide adequate information on all pedestrian infrastructure features. There is an endless need for more and new annotated data for pedestrian infrastructure objects. Luckily, foundation models that can generalize to data that they have not seen are considerably successful in tasks including natural language process (NLP) and computer vision \cite{bommasani2021opportunities}. The Segment Anything Model (SAM) \cite{kirillov2023segment}, released by Meta, is one of the foundation models for image segmentation, showing remarkable performance in zero-shot learning. In the context of pedestrian infrastructure inventory, SAM has great, but unexplored potentials that could significantly reduce the need for labeling a large amount of data and therefore improve the generalizability of trained computer vision models. 

In this project, we aim to develop a novel approach that can process multi-mode geospatial datasets, including mobile LiDAR and satellite imagery, by using the Segment Anything Model (SAM), to generate comprehensive \textit {\textbf{pedestrian accessible infrastructure}} databases for all pedestrian types. Throughout our project, we intend to address two entwined questions: 
\begin{enumerate}[label=(\arabic*)]
    \item   How well does SAM generalize to mobile LiDAR and satellite imagery data in representative segmentation tasks for pedestrian accessible infrastructure inventory work?
    \item   What data representations can effectively boost zero-shot image segmentation performance with SAM for pedestrian  accessible infrastructure inventory work?
\end{enumerate}

The pedestrian accessible infrastructure features considered in this paper are not only inclusive of \textbf{traditional features} such as \textit{sidewalks}, \textit{crosswalks}, \textit{detectable warning surfaces}, and \textit{curb ramps}, but are also inclusive of numerous types of \textbf{street furniture}, particularly those impacting walkability and accessibility. These include but are not limited to: \textit{benches; bollards; fire hydrants; landscapes; mailboxes; manhole covers; memorials; phone booths; parking meters; posts; public sculptures; public vending machines; stairs; storm water inlets; traffic barriers; trees;} and \textit{waste containers}. Each of the objects mentioned above certainly has the potential to hinder navigation as they can occupy space on the sidewalks and pose as obstacles that can be particularly treacherous for the disabled. However, through the utilization of their locations on the sidewalks, the objects can also assist in navigation by serving as landmarks and indicators for walkers. 

\section{Literature Review}
As an indispensable infrastructure, sidewalks support the essential daily trips of pedestrians and especially those with travel-limiting disabilities. Developing a comprehensive sidewalk inventory attracts increasing interests among researchers from various fields, such as urban studies and urban mobility. Based on the data sources, we classified sidewalk-related studies into three categories: (1) image-based, (2) point cloud-based, and (3) data fusion-based. 

\subsection{Image-based}
Most image-based studies focus on localizing the sidewalks using satellite imagery or geo-tagged street level images. Luo, Wu \cite{luo2019developing} extracted connected sidewalk network from aerial images and the common occlusion problem resulted from trees and their shadows can bring large uncertainty to the detected sidewalks under the Bird’s Eye View \cite{senlet2012segmentation}. Hosseini et al. \cite{hosseini2023mapping} developed a scalable satellite imagery based sidewalk inventory method. Street view images are another data sources used for assessing sidewalk accessibilities. Rao et al. \cite{rao2016vision} developed computer vision methods to detect potholes and uneven surfaces on sidewalks for assisting blind people. Li et al. \cite{li2017presight} developed Presight, a system capable of detecting in real-time sidewalk accessibility problems during trips. Many recent studies also explored the use of large open geospatial data sets, particularly Google Street View (GSV), to create sidewalk accessibility databases \cite{kang2021developing, hara2014tohme}. A common issue about these image based methods is that detailed geometry attributes of the sidewalks, such as grades and cross slopes, are beyond the scope of 2D optical images considering their limited spatial accuracy. But these geometric features are essential to accessibility analysis for those who are visually impaired or who rely heavily on wheelchairs, canes, crutches, or walkers \cite{schwartz2021human}. 

\subsection{Point Cloud-based}
Recently, Light Detection and Ranging (LiDAR) technology has been increasingly used for creating infrastructure inventory. Horváth, Pozna \cite{horvath2021real} detected sidewalk edge in real-time using point cloud collected by autonomous driving cars. Hou and Ai \cite{hou2020network} extracted sidewalks through segmenting point clouds using deep neural networks. Esmor et al. \cite{esmoris2023characterizing} used mobile LiDAR data to characterize zebra cross zones. But these typical supervised learning-based 3D segmentation requires large annotated training data sets, and creating these data sets are highly time-consuming and labor-intensive. 

\subsection{Data Fusion-based}
Recognizing the limitations of a single type of data collection method, researchers have also studied the use of multi-modal data to improve the model performance. Luaces et al. \cite{ai2016automated} studied the fusion of LiDAR data and social media data for sidewalk inventory.  Ai and Tsai \cite{ai2016automated} extracted sidewalks and curb ramps using both point clouds and video frames. However, in most data fusion based research, some factors, such as the misalignment of different sensors, can adversely affect the results of multi-modal fusion.\\

Scaling up training database to include more categories of pedestrian infrastructure objects, such as those described in the introduction section is a common challenge in existing pedestrian infrastructure inventory research. Designing a framework with strong generalization ability has been a long-sought research goal. The Segment Anything Model (SAM) is a foundation model for image segmentation tasks \cite{kirillov2023segment}. SAM’s support with zero-shot image segmentation with various prompts such as points, boxes, and masks make it highly attractive for image segmentation tasks in pedestrian infrastructure inventory work. However, there is currently no assessment of SAM based zero-shot segmentation on data used in pedestrian infrastructure inventory. This leads to the present project, which aims to evaluate the performance of SAM in segmenting a broad range of pedestrian infrastructure elements from multi-sourced geospatial data sets. 
\section{Methodology}
\subsection{Description of the Data}

Satellite imagery and 3D point cloud data are among common types of data that have been studied for extracting pedestrian infrastructure features at the network level. In our project, we use mobile LiDAR point cloud data as the primary data source, and bring in Google satellite imagery as the secondary data source. In the XYZ coordinate system of the LiDAR data, where z is the direction of the elevation, the x and y extents (on the ground) of each mobile LiDAR data file are used to fetch Google satellite imagery. In this way, the mobile LiDAR point cloud data and satellite imagery are geo-referenced and aligned so that pedestrian infrastructure features exacted from each data source can be combined within the same geospatial coordinate system. 

\begin{figure}[H]
\centering
\includegraphics[width=10 cm]{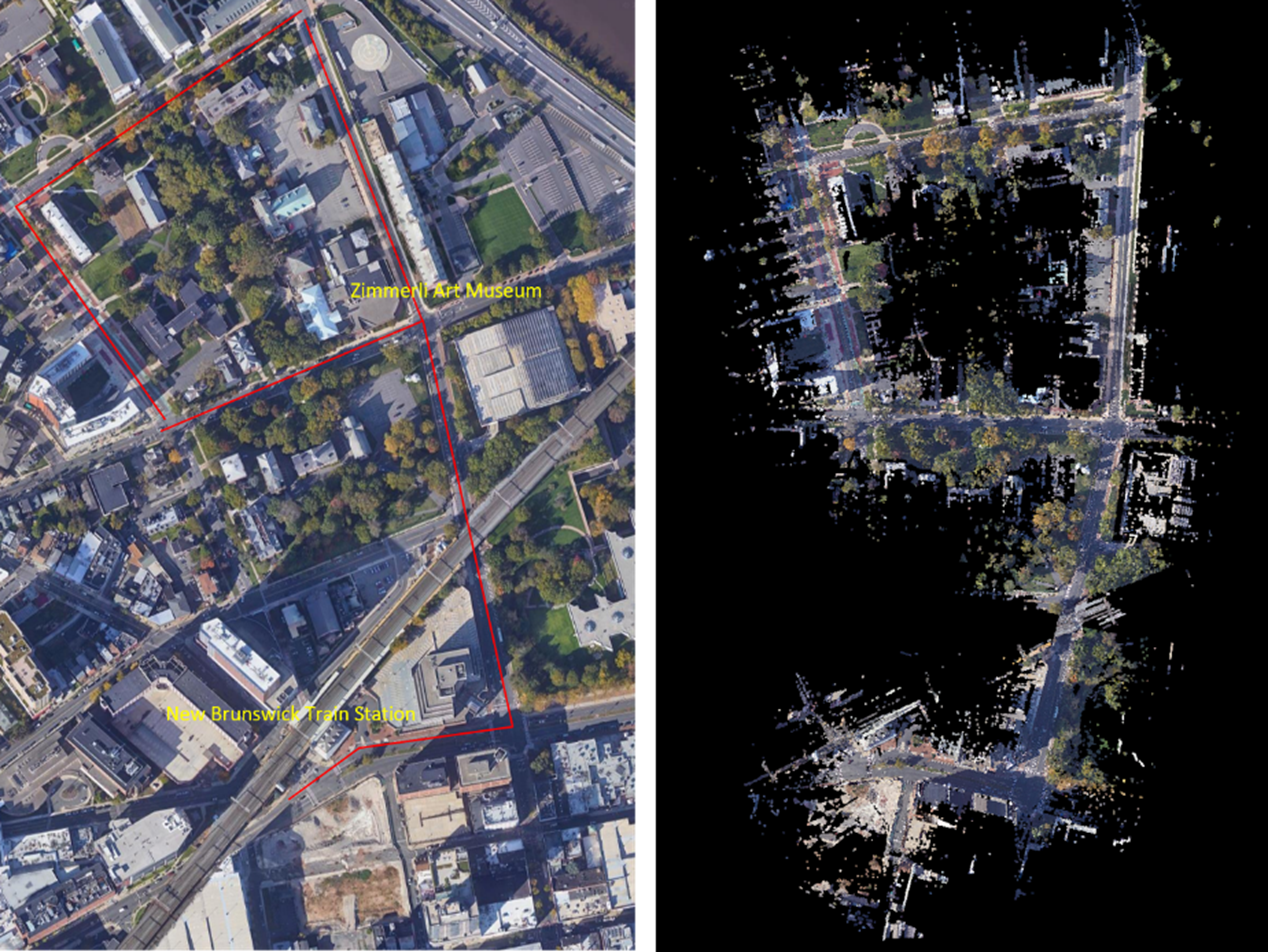}
\caption{Evaluation Data Set (left: Google satellite imagery and right: the mobile LiDAR data) along a route in downtown New Brunswick, NJ}
\label{fig1}
\end{figure}

We use the mobile LiDAR data collected in downtown of New Brunswick, New Jersey and the relevant satellite imagery to evaluate various data representations and the SAM-based pedestrian infrastructure extraction approach. The pedestrian infrastructure of our interest is those along a route connecting the New Brunswick Train Station to the Rutgers Zimmerli Art Museum (Figure \ref{fig1}). The data for this route was recently captured with a survey-grade mobile mapping system with 1MHz laser pulse rate. The system comprises an Applanix 420 navigation system, a ZF 9012 laser profiler, and a ZF map cam system. The point cloud data are divided into eight individual point cloud files, with each file having a size of 500MB.  The route features multiple transportation modes in automobiles, bikes, e-scooters, and pedestrians as well as complex intersection designs. Along the routes there are Rutgers campus buildings, business establishments, hospitals, and many other types of amenities. The Zimmerli Art Museum also often hosts special events for people with disabilities such as wheelchair users and people with visual impairment. All these facilities and activities lead to heavy uses of pedestrian infrastructures along this route by all types of pedestrians. The work described here directly contributes to the development of digital twins of cities.

\unskip

\begin{figure}[H]
\begin{adjustwidth}{-\extralength}{0cm}
\centering
\captionsetup{justification=centering}
\includegraphics[width=15.5cm]{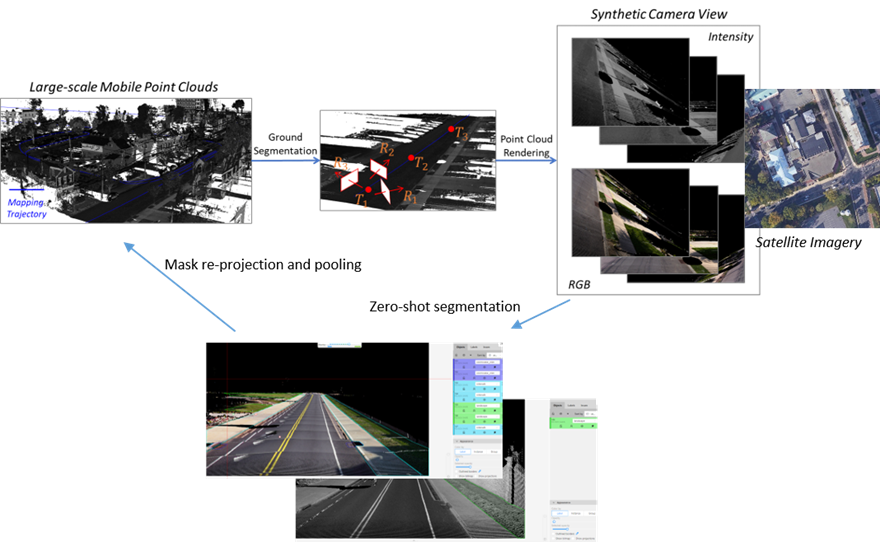}
\end{adjustwidth}
\caption{Elements of the proposed SAM-based multi-sourced pedestrian accessible infrastructure segmentation workflow.}
\label{fig2}
\end{figure}  

\subsection{SAM-based Workflow}
The SAM-based multi-sourced pedestrian accessible infrastructure segmentation workflow includes the following three steps: (1) LiDAR data pre-processing; (2) Data projection and alignment; (3) Zero-shot segmentation; and (4) Re-projection and pooling of segmentation masks. Figure \ref{fig2} illustrates the process. Large-scale mobile point cloud data is segmented in the first step into ground and non-ground classes. Two different synthetic camera views are generated and aligned with the corresponding satellite image in the second step. In the third step, zero-shot segmentation is performed on the set of 2D images. Then in the fourth step, object masks generated from the segmentation are reprojected back to the 3D geospatial space.

\subsubsection{Data pre-processing }
The LiDAR point cloud data sets were pre-processed with a common classification approach  which involves classifying the point cloud into classes including ground, low vegetation, medium vegetation, and high vegetation. This approach starts with classifying ground points with the ground segmentation algorithms, namely Simple Morphological Filter (SMRF) \cite{schwartz2021human}. Once the ground points are classified, heights from ground are used to classify low vegetation, medium vegetation, and high vegetation. 

\subsubsection{Data projection and alignment}
\begin{figure}[H]
\centering
\includegraphics[width=14 cm]{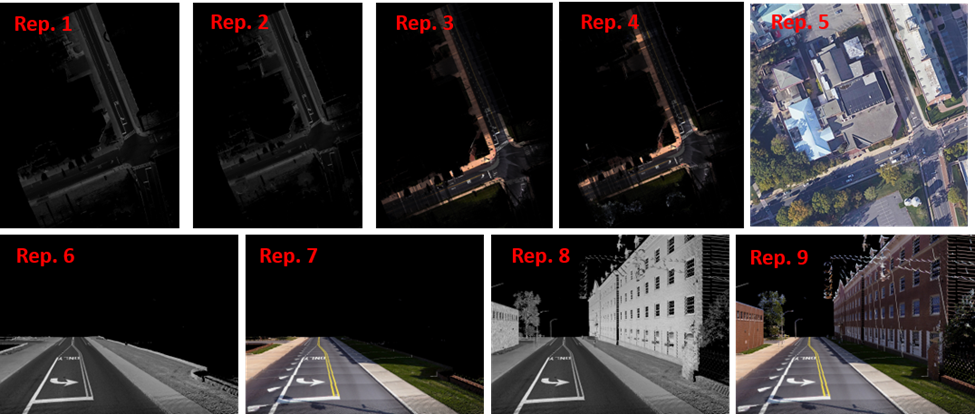}
\caption{Image Representations: Rep. 1 – BEV with only ground class and intensity values, Rep. 2 – BEV with all classes and intensity values, Rep. 3 – BEV with only ground class and color values, Rep. 4 – BEV with all classes and color values, Rep. 5 – Satellite imagery, Rep. 6 – Street view with only ground class and intensity values, Rep. 7 – Street view with only ground class and color values, Rep. 8 – Street view with all classes and intensity values, and Rep. 9 – Street view with all classes and color values.}
\label{fig3}
\end{figure}

To work with the SAM-based zero-shot learning, we created a stack of 2D images from mobile LiDAR data by projecting 3D point cloud data onto different planes. The first type of generated 2D images are Bird Eye View (BEV) images, also called vertical ortho-rectified images, by projecting point clouds onto a horizontal grid with a grid size of $0.1 feet \times 0.1 feet$. The choice of this grid size consummates with the resolution of the point cloud data sets. Four (4) variants of BEV images are generated by changing both the class of included points and the pixel attributes (Figure \ref{fig3}, Rep. 1 to Rep 4). The possible values for the class of included points are with all classes and with only ground and low vegetation classes. The possible attributes for the pixel values are laser intensity and color. The color value of a laser point is obtained by projecting the point onto color images which were also collected along with point cloud data. These BEV images are essentially the commonly used GeoTIFF files that can be directly consumed by many GIS programs such as ESRI ArcGIS, the most popular GIS software. They also provide a convenient way to retrieve satellite imagery of the same data extent through the WMS service from Google. The mobile LiDAR point cloud data and satellite imagery are geo-referenced and aligned so that pedestrian infrastructure features exacted from each data source can be combined within a same geospatial coordinate system. A geo-referenced satellite image is shown as Rep. 5 in Figure \ref{fig3}.

The second type of generated 2D images are street view images. These images are obtained by simulating virtual cameras and capturing snapshots at street level within the scenes represented by a mobile LiDAR point cloud data file (Figure \ref{fig2}). Since the parameters of these virtual cameras are known, every pixel in the generated images can be back-traced to 3D points in the original point clouds. The details of the synthetic image generation process is provided in Section “Synthesis image generation details”. Similar to BEV images, four variants of street images for each snapshot are created by varying the class of points included and the pixel values (Figure \ref{fig3}, Rep. 6 to Rep. 9). 

\subsubsection{Zero-shot segmentation}
Meta provides a SAM demo website that allows users to upload images to try zero-shot segmentation. But this website is not scalable to image segmentation experiments involving many images. Instead, we deployed a SAM-based annotation project on CVAT, a web-based image annotation platform. In this web-based project, users can fire the request to use SAM as an interactor to do zero-shot segmentation on images. Basically, the SAM predicts object masks given prompts (point clicks) that indicate the desired object. A list  of pedestrian infrastructure features is provided in Table \ref{tab:1} below to allow users to choose which pedestrian infrastructure object to segment. The object masks are archived and can be exported in common image annotation formats such as COCO 1.0 \cite{COCO}. 


\subsubsection{{Mask reprojection and pooling}}
In this project, the object masks are essentially polygons enclosing the boundaries of specific pedestrian infrastructure objects. The polygons are described in the image pixel coordinate system. Once zero-shot segmentation tasks are completed on a stack of images for the same area, we need to reproject the object masks back to the 3D geospatial space. This reprojection also enables the pooling of masks generated from multiple images to segment the original point cloud, allowing detailed geometric analysis, such as slope and grade assessment, on specific objects. Furthermore, it will facilitate the assembly of large 2D and 3D GIS maps. 

\subsection{Synthesis Image Generation Details}
Given the mapping trajectory, the positions distancing equally sampled along the trajectory are used as the centers of the virtual camera locations. The focal length $f$ and pixel size $ps$ of the virtual camera are set as $4.15mm$ and $1.22\mu m$ and the camera matrix $A$ of the virtual camera with a view size of $4032 \times 3024$ $(width \times height)$ is defined in equation (\ref{eq:1}) below: 
\begin{linenomath}
\begin{equation}
\mathbf{A} = 
\begin{bmatrix}
    \frac{f}{ps} & 0 & \frac{width}{2} \\
    0 & \frac{f}{ps} & \frac{height}{2} \\
    0 & 0 & 1 \\
\end{bmatrix}
\label{eq:1}
\end{equation}
\end{linenomath}

The distance between two camera locations should ensure that overlap exists between the synthetic views to cover all the ground points. For each position of the camera, $T$, an arbitrary camera orientation $R$ can be defined; one front view and two side views are set to cover both the road and sidewalks in this study. The joint rotation-translation matrix $R_t$ can be calculated using the camera position vector and orientation matrix. With customized camera matrix $A$ and rotation-translation matrix $R_t$, virtual camera views are generated using the Open3D library based on equation (2) below [26]: 
\begin{linenomath}
\begin{equation}
\mathbf{s} \begin{bmatrix} u \\ v \\ 1 \end{bmatrix} = \mathbf{A} \mathbf{R}_t \begin{bmatrix} X \\ Y \\ Z \\ 1 \end{bmatrix}
\end{equation}
\end{linenomath}

The parameter “point size” is a critical factor which affects the final rendering results. We set point size as 8 in the Open3D library. Both intensity-based views and RGB-based views are generated to provide rich texture information for sidewalk-related infrastructure extraction. With the segmentation mask on the synthetic camera views, we can project the 2D mask back to the 3D space with the inverse operation of equation (2).

Figure \ref{fig3} shows examples of a full stack of images generated from a sample point cloud file, plus a satellite image (Rep. 5) of the same scene. These images share the same geospatial coordinate system, and all the pixels in the images derived from the point cloud file have 3D coordinates. In short, every segmentation mask generated by segmentation tools such as SAM on these images can be projected onto a planimetric map to create a comprehensive inventory of pedestrian infrastructure features.  

\subsection{Experiment and Assessment}

One of the goals of this research is to identify data representations that can facilitate zero-shot segmentation with SAM. We evaluated the level of SAM-based pedestrian infrastructure extraction, segmentation of these infrastructure elements, from these images of different representations and perspectives. A total of eight sets of images are used to test the performance of SAM with zero-shot learning. Each set of images includes 4 BEV images and 24 street view images generated from mobile LiDAR point cloud, plus 1 satellite image. This brings the total number of images to 232. 

The choice of parameters such as included classes of point clouds and types of pixel values is driven by two considerations. First, including some classes of point clouds may present challenges rather than assistance in extracting pedestrian infrastructures. For example, points belong to tree classes may occlude sidewalks. Second, reflectance in different light spectrums are features useful to identify different types of pedestrian infrastructures. For example, color values are good for sidewalk segmentation, but intensity values are good for crosswalk detection. 

The levels of zero-shot segmentation are expressed with no extraction (N), partial extraction (P), and complete extraction (C). If a particular type of pedestrian infrastructure object does not exist in the data set, a N/A was recorded. We also provide the reasons behind partial and no extraction in the hope to provide insights into why the extraction failed, and more importantly on the possible ways of improving extractions with data fusion.

\begin{table}[H]
\caption{Evaluation results of different pedestrian infrastructure features using the SAM model. N: no extraction, P: partial extraction, C: complete extraction and N/A: infrastructure type unavailable.\label{tab:1}}
	\begin{adjustwidth}{-\extralength}{0cm}
		\newcolumntype{C}{>{\centering\arraybackslash}X}
		\begin{tabularx}{\fulllength}{l|l|CCCCCCCCC}
			\toprule
			& \textbf{Pedestrian Infrastructure Features} & \textbf{Rep. 1} & \textbf{Rep. 2} & \textbf{Rep. 3} & \textbf{Rep. 4} & \textbf{Rep. 5} & \textbf{Rep. 6} & \textbf{Rep. 7} & \textbf{Rep. 8} & \textbf{Rep. 9} \\
			\midrule
			\multirow{9}{*}{\textit{Planimetric features}} & Sidewalk & P & P & P & P & P & C & C & C & C \\
			& Crosswalk & C & C & C & C & C & C & C & C & C \\
			& Curb ramp & P & P & P & P & N & C & C & C & C \\
			& Landscape & P & P & P & P & P & C & C & P & P \\
			& Stair & N & N & P & P & N & N & N & C & C \\
			& Detectable warning surface & N & C & N & C & N & C & C & C & C \\
   		& Storm water inlet & N & N & N & N & N & C & C & C & C \\
      	& Manhole cover & N & N & N & N & N & C & C & C & C \\
			& Traffic barrier & N & N & N & N & N & N & N & C & C \\
			& Retaining wall & N & N & N & N & N & C & C & C & C \\
			\midrule
			\multirow{14}{*}{\textit{Volumetric features}} & Bench & N/A & N/A & N/A & N/A & N/A & N/A & N/A & N/A & N/A \\
			& Bollard & N/A & N/A & N/A & N/A & N/A & N/A & N/A & N/A & N/A \\
			& Fire hydrant & N & N & N & N & N & N & N & C & C \\
			& Mailbox & N/A & N/A & N/A & N/A & N/A & N/A & N/A & N/A & N/A \\
			& Memorial & N/A & N/A & N/A & N/A & N/A & N/A & N/A & N/A & N/A \\
			& Phone booth & N/A & N/A & N/A & N/A & N/A & N/A & N/A & N/A & N/A \\
			& Parking meter & N & N & N & N & N & C & C & C & C \\
			& Post & N & N & N & N & N & N & N & C & C \\
			& Public Sculpture & N/A & N/A & N/A & N/A & N/A & N/A & N/A & N/A & N/A \\
			& Public vending machine & N/A & N/A & N/A & N/A & N/A & N/A & N/A & N/A & N/A \\
			& Tree trunk & N & N & N & N & N & N & N & C & C \\
   		& Tree Canopy & N & N & P & P & P & N & N & P & P \\
			& Waste container & N & N & N & N & N & N & N & C & C \\
			\bottomrule
		\end{tabularx}
	\end{adjustwidth}
\end{table}

\section{Results and Discussion}

Table \ref{tab:1} gives a snapshot of the evaluation results. The findings can be summarized as: 
\begin{enumerate}[label=(\arabic*)]
    \item   \textit{\textbf{Occlusions.}} Sidewalks can be severely occluded by tree canopies in satellite imagery, leading to incomplete segmentation of sidewalks. Occlusions to sidewalks by vehicles and other street objects are much less a concern when satellite imagery is used, but they limit what can be extracted from mobile LiDAR point clouds on busy street environments. On the other hand, mobile LiDAR data can well capture data about sidewalk under tree canopies. Therefore, combining mobile LiDAR data with satellite imagery can address each method’s weakness and allow for complete extraction of sidewalks with SAM in most cases (Figure \ref{fig4}). However, we have found a number of pitfalls in using SAM. SAM-based zero-shot learning on our images was not stable as SAM crashed multiple times.  A considerable number of prompts, clicks on areas in need of segmentation, are needed in order to achieve good segmentation. This is likely due to the size and resolution of these images, which are as large as $7000 \times 10000$.  In other words, SAM can work more effectively with satellite and BEV images with smaller fields of view or lower image resolutions. However, analyzing satellite imagery or BEV images at smaller patches may result in impractical larger number of interactions and/or lower performance in segmentation.
    \begin{figure}[H]
    \centering
    \includegraphics[width=12 cm]{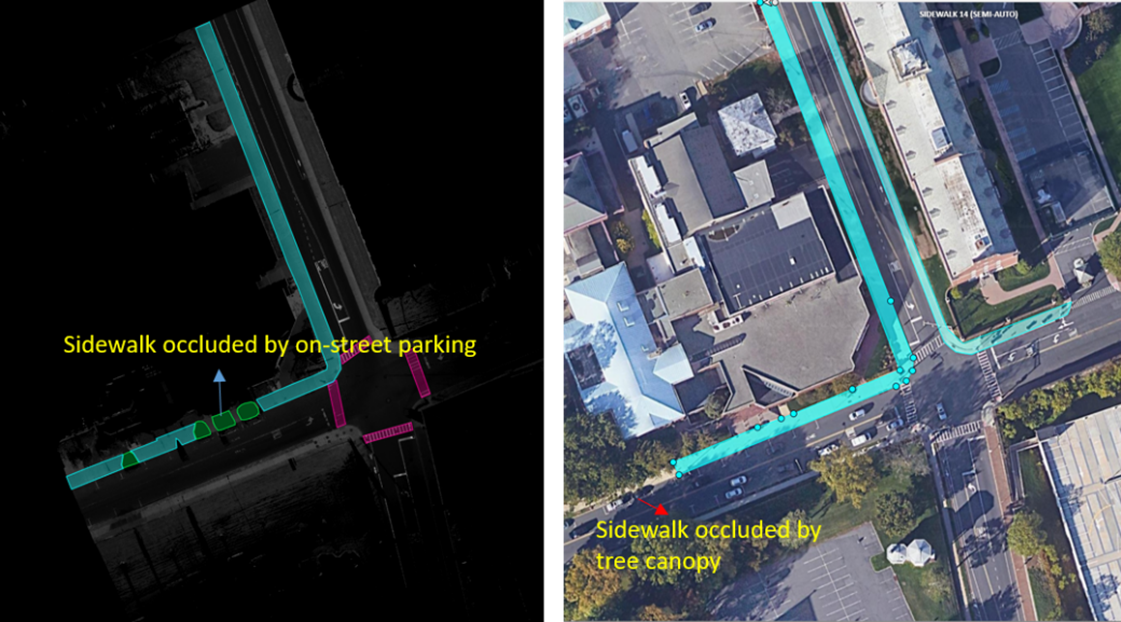}
    \caption{Sidewalk extraction with the fusion of mobile LiDAR and satellite imagery. The left image shows that the sidewalk occluded by on-street parking in the LiDAR data can be extracted in satellite imagery, while the right image shows that the sidewalk occluded by tree canopy in the satellite imagery can be extracted in the LiDAR data.}
    \label{fig4}
    \end{figure}

    \item  \textit{\textbf{Small volumetric features.}} It is generally difficult to extract small volumetric pedestrian infrastructure features, such as fire hydrants, parking meters, and posts, as well as small planimetric features such as manhole covers  from either satellite imagery or BEV images generated from mobile LiDAR data. But they can be identified and segmented in high quality with SAM with very few prompts from street view images that are generated from mobile LiDAR point cloud data (Figure \ref{fig5}). 
    \begin{figure}[H]
    \centering
    \includegraphics[width=14 cm]{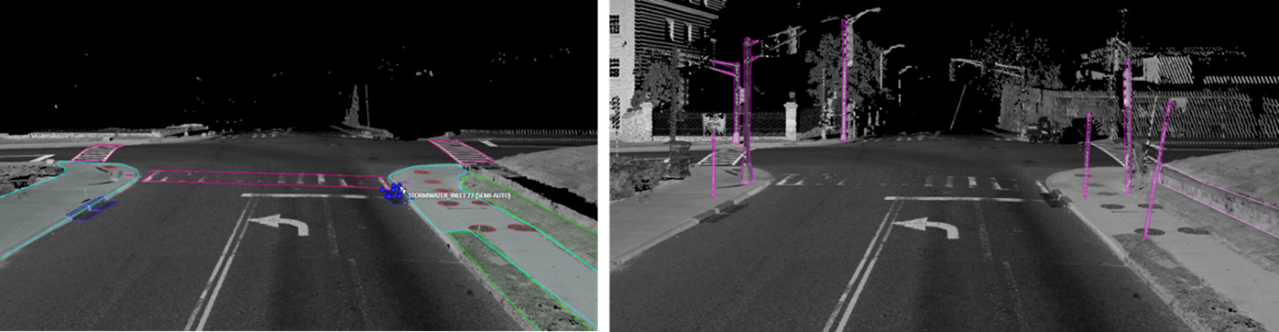}
    \caption{Segmentation of planimetric (left) and volumetric pedestrian infrastructure features (right) on street view images generated from point cloud data.}
    \label{fig5}
    \end{figure}

    \item  \textit{\textbf{Planimetric features.}} The SAM models achieved remarkable performance in segmenting planimetric pedestrian infrastructure features such as sidewalk, crosswalk, detectable warning surfaces, and curb ramps under zero-shot learning scenarios on street view images that are generated from mobile LiDAR point cloud with only ground and low vegetation classes (Figure \ref{fig5}). Very few prompts are needed to achieve complete segmentation of these features. The exception cases are streets with significant amounts of street-level occlusions such as on-street parking. On the other hand, over-segmentation of planimetric pedestrian infrastructure features can happen frequently in street view images generated from mobile LiDAR point cloud with all classes (Figure \ref{fig6}). 
    \begin{figure}[H]
    \centering
    \includegraphics[width=12 cm]{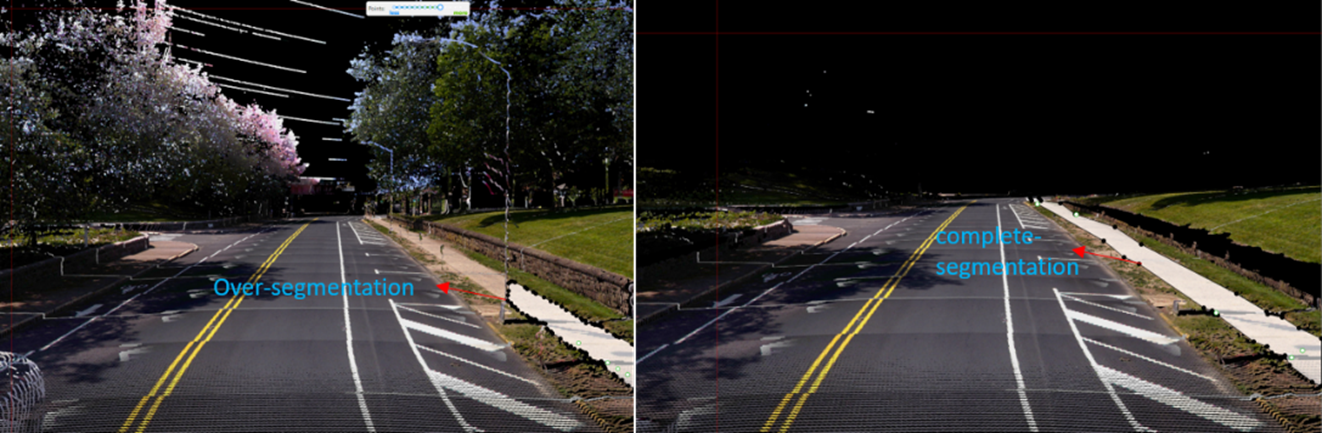}
    \caption{Over-segmentation on street view images with all classes}
    \label{fig6}
    \end{figure}

    \item \textit{\textbf{Color versus reflectance images.}}  Segmenting pedestrian infrastructure objects in BEV and street-level images with color pixel values in some cases is easier than that of those intensity-based images as the color provides stronger clues about the boundary of these objects in the images. But this is only true in cases where the shadow is not of a concern. The intensity values, representing the strength of laser reflectance, are not influenced by shadows and light conditions. They are consistent for the same types of materials. In addition, the color information in Rep. 3, 4, 7, and 9 in Figure \ref{fig3} are the results of data fusion, which involves aligning outputs of the laser scanners and cameras while the platform hosting them is moving at 10-40 miles per hour. The error in alignment is inevitable, but will have significant impacts on the quality of data fusion. 

    \item  \textit{\textbf{Street view images.}} Street-level images with intensity values, generated from mobile LiDAR data, have the overall best performance in working with SAM to segment both planimetric and volumetric pedestrian infrastructure features. But if street-level images, those generated from mobile LiDAR data, are used as the sole source of data for extracting pedestrian infrastructure objects, we must address the street-level occlusion problem. One  way to alleviate the problem is to conduct nighttime data collection as laser scanning alone is not impacted by daylight and there will be far less occlusions due to vehicles at night. In addition, this can be possibly done through automated recognition of presences of street occlusion objects and automated filling of occluded areas. 
    \begin{figure}[H]
    \centering
    \includegraphics[width=12 cm]{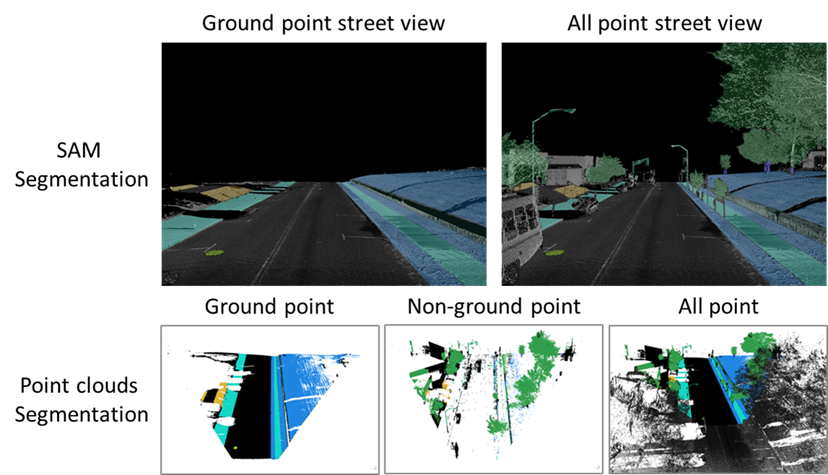}
    \caption{Example results of SAM-based workflow on mobile LiDAR point cloud data.}
    \label{fig7}
    \end{figure}
      
\end{enumerate}  
    
\begin{figure}[H]
    \centering
   \includegraphics[width=12 cm]{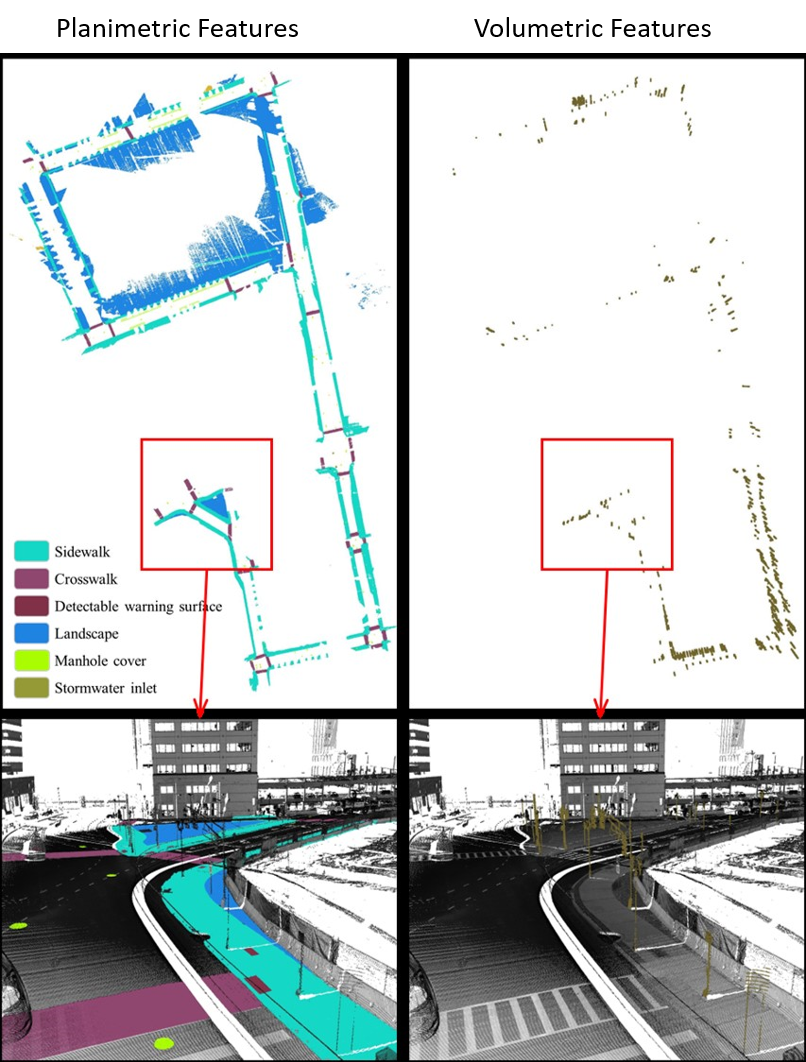}
    \caption{3D point cloud semantic segmentation results. The left ground-based points mainly include sidewalks, crosswalks, detectable warning surfaces, landscape, manhole cover, and stormwater inlet. The left pictures show the volumetric features on the sidewalks, such as posts.}
    \label{fig9}
\end{figure}

\begin{figure}[H]
    \centering
    \includegraphics[width=14 cm]{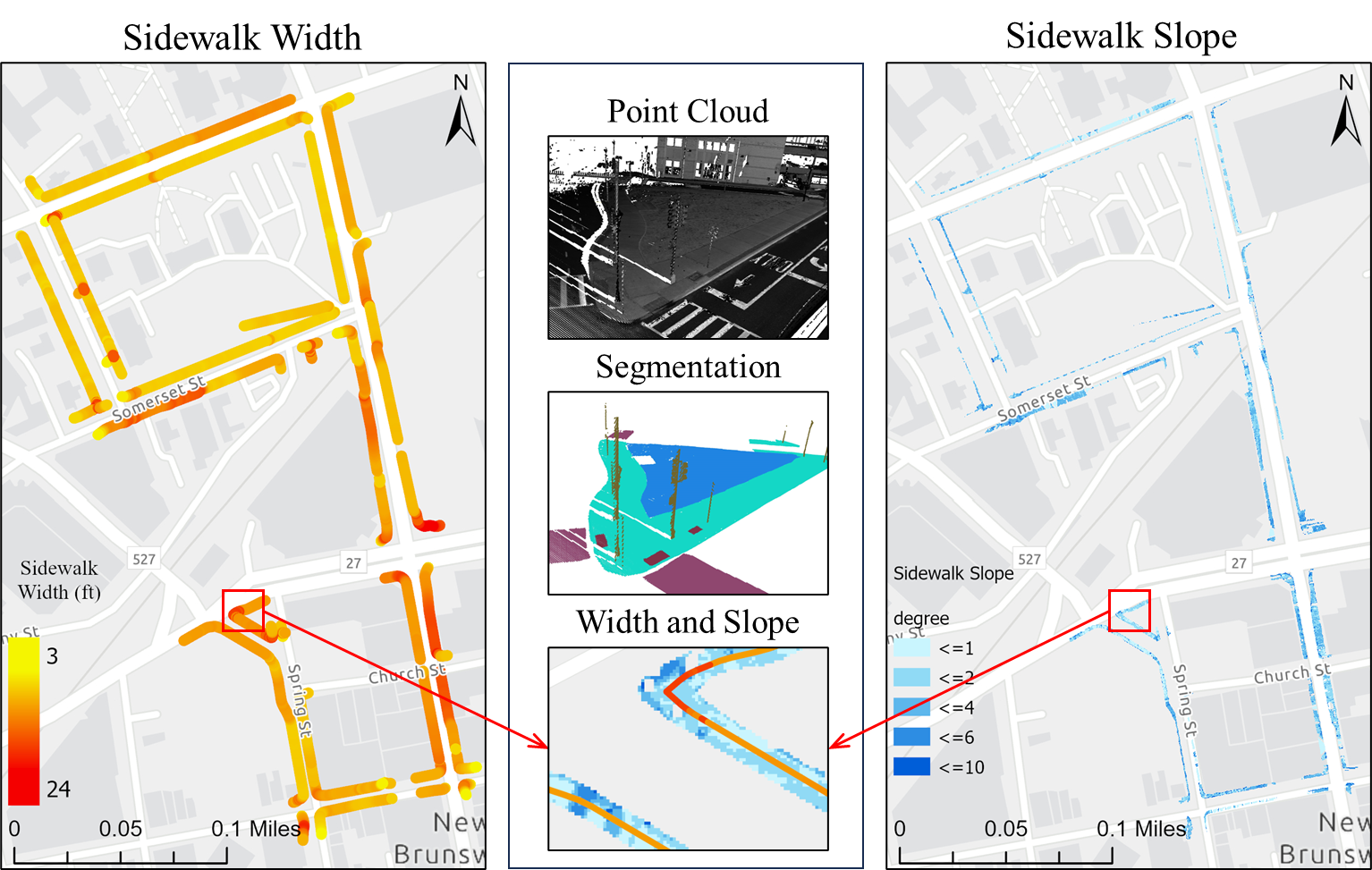}
    \caption{Computed widths and slopes for the entire network of sidewalk in the study area.}
    \label{fig10}
\end{figure}

    Overall, the SAM-based workflow is capable of providing an end-to-end solution for GIS professionals to digitize pedestrian infrastructure objects with high accuracy. This is particularly true when the right image data representation is used (Figure \ref{fig7}). As part of the assessment, we did a complete extraction of the sidewalk features in the study area from the street views (Figure \ref{fig9}). These street views are intensity images from two types of point clouds, one with only the ground class and the other type with all the classes. Beyond showing the 2D locations of these pedestrian infrastructures, we further processed the segmented point clouds to characterize the sidewalk with important parameters including width and slope (Figure \ref{fig10}). These results showed that the workflow by itself can already serve as a powerful tool for GIS professionals to process mobile LiDAR data into actionable information for improving pedestrian infrastructure. 

\section{Conclusion}
In this paper, we designed and optimized a SAM-based pedestrian infrastructure segmentation workflow capable of efficiently processing multi-sourced geospatial data. We used an expanded definition of pedestrian infrastructure, termed as \textit{pedestrian accessible infrastructure}, which includes street furniture objects often omitted from the traditional definition. Our contributions lie in producing the necessary knowledge to answer the following two questions for pestrian accessibility. First, which data representation can facilitate zero-shot segmentation of infrastructure objects with SAM? Second, how well does the SAM-based method perform in segmenting pedestrian infrastructure objects? Our findings indicate that street view images generated from mobile LiDAR point cloud data, when paired with satellite imagery data, can work efficiently with SAM to create a scalable pedestrian accessible infrastructure inventory approach with immediate benefits to GIS professionals, transportation owners and walkers especially those with travel-limiting disabilities. We demonstrated that the SAM-based workflow is capable of extracting sidewalk widths and slopes from the street view images generated from mobile LiDAR data. This information is essential for individuals with visual and mobility impairments. 

We would like to note that the scope of this paper is mostly limited to locating basic pedestrian accessible infrastructure features. Additional processing is often needed to truly assess the compliance of these infrastructure features, especially those more complicated ones such as curb ramps. These additional processing steps will be included in our subsequent research. Our future research will also incorporate the training of deep learning models based on the annotated pedestrian infrastructure data to fully automate the entire workflow.

\section{Acknowledgement}
This work was supported by the National Science Foundation (NSF) through Awards $\#2131186$ (CISE-MSI), $\#1951890$ (S\&CC), $\#1827505$ (PFI) and $\#1737533$ (S\&CC). The work is also partially supported by New Jersey Department of Transportation and Federal Highway Administration Research Project 21-60168, Middlesex County Resolution 21-821-R, the US Air Force Office of Scientific Research (AFOSR) via Award $\#FA9550-21-1-0082$ and the ODNI Intelligence Community Center for Academic Excellence (IC CAE) at Rutgers University ($\#HHM402-19-1-0003$ and $\#HHM402-18-1-0007$). 

\begin{adjustwidth}{-\extralength}{0cm}

\reftitle{References}

{\small
\bibliographystyle{ieeetr}
\bibliography{ref}
}

\PublishersNote{}
\end{adjustwidth}
\end{document}